\documentclass[showpacs,preprint,amssymb]{revtex4}
\usepackage{graphicx}
\usepackage{dcolumn}
\usepackage{slashed}

\usepackage{bm}
\def\be{\begin{equation}} 
\def\ee{\end{equation}}
\def\bea{\begin{eqnarray}} 
\def\eea{\end{eqnarray}}
\def\line{\hbox to \hsize}    
\def\frac #1#2{{#1\over #2}}

\def\tr{{\rm  tr\,}}
\def\det{{\rm det\,}}

\def\bz{{\bar z}}

\def\det{{\rm det\,}}

\def\1{\mbox{\bf 1}}
\def\bm#1{\mbox{\boldmath$#1$}} 

\begin{document}

\title{Mixed Anomalies: Chiral Vortical Effect and   the Sommerfeld Expansion }

\author{ Michael Stone}

\affiliation{University of Illinois, Department of Physics\\ 1110 W. Green St.\\
Urbana, IL 61801 USA\\E-mail: m-stone5@illinois.edu}   

\author{ JiYoung Kim}

\affiliation{University of Illinois, Department of Physics\\ 1110 W. Green St.\\
Urbana, IL 61801 USA\\E-mail: jkim623@illinois.edu}

\begin{abstract}  We discuss the connection between the integer moments of the Fermi distribution function that occur in the Sommerfeld expansion and the coefficients  that occur in anomalous conservation laws for chiral fermions. As an illustration  we extract  the chiral magneto-thermal energy current from the mixed gauge-gravity anomaly in the 3+1 dimensional energy-momentum conservation law. We then use  a similar method to   confirm the conjecture  that the   $T^2/12$  thermal contribution to the chiral vortical effect (CVE) current arises from  the gravitational Pontryagin  term in the 3+1-dimensional chiral anomaly.   

\end{abstract}


\maketitle

\section{Introduction}

A recent experiment \cite{gooth} and its widespread coverage in the media \cite{changNYT} have focussed attention on the idea that the physics of a system containing chiral fermions can be influenced by effects of gravitational origin even in flat space-time \cite{landsteiner}.  These effects occur because the coefficients in certain  constitutive relations for transport currents  are related to the coefficients in corresponding anomalous conservation laws.  As anomalies are not renormalized by interactions, these anomaly-induced, non-dissipative,  contributions to transport currents  should take the same values in both strongly-coupled and free theories. In the free  case  the currents can  be computed without reference to any  anomaly,  and  the free-theory computations reduce to the evaluation  of  integer moments  of the Fermi function  that   turn out to  be polynomials in the temperature and chemical potential.  One is left with a sense that  these Sommerfeld-expansion  integrals somehow know about  anomalies.  This  impression   was made concrete  by  Loganayagam and   Sur{\'o}wka \cite{ideal} who observed that a  generating function for the integer moments of the 
Fermi function  bears a close resemblance to the the product of the A-roof genus and   the total  Chern character  that occurs in the general-dimension  Dirac index theorem --- and which,  {\it via\/} the Bardeen-Zumino descent equations \cite{bardeen-zumino}, is the ultimate source of the anomalies. Their observation lead them to a ``replacement rule'' that allowed them    to compute anomaly-induced contributions to transport and fluid dynamics in $N$ space-time  dimensions directly from the anomaly polynomial   in $N+2$ dimensions \cite{ideal,azeyanagi,banarjee-loga,jensen1,jensen2,jensen3,dutta1,stone-dwivedi}.   

In this paper we  illustrate some of  ideas  by  computing two of  these currents --- the thermo-magnetic current that plays a central role in \cite{gooth},  and the thermal contribution to the chiral vortical effect (CVE) current that arises when a chiral fermion is in thermal equilibrium in a rotating frame --- both from the free theory and from the corresponding anomaly.  The first example is merely a  repackaging of the gravitational-anomaly derivation of Hawking radiation \cite{wilczek1,wilczek2,banerjee}  but it serves to set the stage for an explicit   confirmation  of the conjecture   \cite{landsteiner} that  the thermal component of the CVE is related to the gravitational anomaly. These two derivations also  help explicate  the geometric origin of the replacement-rule mapping that takes the first Pontryagin class of the Riemann curvature tensor to minus the square of the temperature.

In section \ref{SEC:Anomalies} we introduce the specific  currents  whose  anomaly-driven origin  we seek to illustrate. In section \ref{SEC:Anomaly-derivations} we will construct   {\it gedanken\/} spacetimes  in which the currents are created  {\it ex nihilo\/}  by  tidal forces in the vicinity of   black-hole event horizons.  In section \ref{SEC:generating} we  use  similarity of the their generating functions and the observation that  often only one of the formal eigenvalues of the curvature tensor will be non-zero to  link the anomalies with  the Sommerfeld integrals. A final section \ref{SEC:discussion} will put these ideas into context.    
 
\section{Anomalies and Anomaly-induced  currents}
\label{SEC:Anomalies}

The ``mixed axial-gravitational anomalies''    that are invoked in  the condensed-matter context in  \cite{gooth} and also \cite{lucas} are  the (3+1)-dimensional anomalous conservation equation     
\be
 \nabla_\mu T^{\mu\nu}= F^{\nu\lambda} J_{N,\lambda} - \frac{1}{384\pi^2}\frac{\epsilon^{\rho\sigma\alpha\beta}}{\sqrt -g }  \nabla_\mu [F_{\rho\sigma} {R^{\nu\mu}}_{\alpha\beta}]
 \label{EQ:anom-em}
 \ee
 for  the    energy-momentum tensor $T^{\mu\nu}$ of a unit-charge right-handed Weyl fermion,
 and the anomalous conservation equation  
 \be
 \nabla_\mu J^\mu_N =- \frac{1} {32 \pi^2} \frac{\epsilon^{\mu\nu\rho\sigma}}{\sqrt -g}  F_{\mu\nu}F_{\rho\sigma}- \frac {1}{768\pi^2}\frac{\epsilon^{\mu\nu\rho\sigma}}{\sqrt -g}{R^\alpha}_{\beta\mu\nu} {R^{\beta}}_{\alpha\rho\sigma}
 \label{EQ:anom-number}
 \ee
for  the  particle-number current $J^\mu_N$.
The right-hand-side of Eq (\ref{EQ:anom-em}) shows that  energy-momentum is delivered  to the fermion from  two sources: the first is  the expected  Lorentz force $F^{\nu\lambda} J_{N,\lambda}$; the second is the gravitational anomaly which requires  a cooperation between  the gauge field $F_{\mu\nu}$ and  the  tidal forces   encoded in the Riemann tensor ${R^{\nu\mu}}_{\alpha\beta}$ of the background space-time geometry.  The  right-hand-side  of  the gauge-current conservation law Eq (\ref{EQ:anom-number})  contains  {\it two\/}  anomalous source terms: a gauge field Chern-character density  and a  geometric   Pontryagin-class density.  The  two equations,  (\ref{EQ:anom-em}) and (\ref{EQ:anom-number}), display   {\it mixed  anomalies\/}  because the   anomalous sources for both the geometry-related energy-momentum tensor $T^{\mu\nu}$    and the gauge-field-related particle-number current $J^\mu_N$ contain expressions  involving the background field that couples to the other.

Anomaly-induced currents appear in  solid-state systems  \cite{gooth,lucas} and also in  relativistic fluid dynamics \cite{neiman-oz} where (in  the  $[-,+,+,+]$ metric convention) 
 we have  \footnote{We are ignoring dissipative effects such as viscosity and diffusion.} 
 \bea
 T^{\mu\nu} &=&   pg^{\mu\nu} +(\epsilon+p) u^\mu u^\nu  + \xi_{TB}(B^\mu u^\nu+ B^\nu u^\mu)+ \xi_{T\omega} (\omega^\mu u^\nu+ \omega^\nu u^\mu), 
 \label{EQ:energy-anomaly}\\
J^\mu_N &= & n u^\mu + \xi_{JB} B^\mu +\xi_{J\omega} \omega^\mu,
\label{EQ:number-anomaly}\\
J^{\mu}_S &=& su^\mu + \xi_{SB} B^\mu +\xi_{S\omega} \omega^\mu.
\label{EQ:entropy-anomaly}
\eea
Here $T^{\mu\nu}$ and $J_N^\mu$ are the energy-momentum tensor and number current  that we have already met, while  $J^\mu_S$ is the entropy current. 
The first terms on the RHS of each of these expressions are the  usual expressions for a relativistic  fluid where  $u^\mu$ denotes  the 4-velocity of the fluid,  and 
 $\epsilon$,   $p$, $n$, and $s$ are   respectively the  energy density, pressure,   particle-number density, and entropy density.  The remaining  anomaly-induced  terms involve  the  angular-velocity   
4-vector  defined by 
\be
 \omega^\mu =  \frac 12  \epsilon^{\mu\nu\sigma\tau} u_\nu \partial_\sigma u_\tau.
 \ee
 With $ \epsilon^{0123}=+1$ and $\epsilon_{0123}=-1$,  and in the  $u^\mu=(1,0,0,0)$ rest frame we have $\omega^\mu= (0,{\bm \Omega})$ where ${\bm \Omega} =\frac 12 \nabla\times {\bf u}$ is the local 3-vector angular velocity. 
The extra  currents also  involve the  magnetic field ${\bf B}$ as it appears to an observer moving at velocity $u^\mu$.  We have 
 \be
 E^\mu = F^{\mu\nu}u_{\nu}, \quad B^\mu = \frac 12 \epsilon^{\mu\nu\sigma\tau} u_\nu F_{\sigma\tau} .  
 \ee
 where $F_{\mu\nu}= \partial_\mu A_\nu-\partial_\nu A_\mu$, $A^\mu= (\phi,{\bf A})$.
Again, in the   $u^\mu=(1,0,0,0)$  rest frame, we have     $E^\mu= (0,{\bf E})$, $B^\mu = (0,{\bf B})$
 and  
\be
\frac 1{4\pi^2} {\bf E}\cdot {\bf B}= \frac 1{4\pi^2} E_\mu B^\mu = -\frac 1{32\pi^2} \epsilon^{\mu\nu\sigma\tau} F_{\mu\nu}F_{\sigma\tau} .
 \label{EQ:anomalyEB}
 \ee
  In relativistic fluid dynamics   the  notion of the ``velocity of the fluid''  requires further specification. We will take $u^\mu$ to be the 4-velocity of the  {\it no-drag} frame introduced in \cite{sadofyev,no-drag}. This is the frame  in which the $\xi$ coefficients  take their simplest form,
  and is  usually  the frame in which  the fluid is in local thermodynamic equilibrium. 

Demanding that no entropy production be  associated with  the anomaly-induced currents  requires \cite{no-drag} that  the  six coefficients $\xi_{TB}$, $\xi_{T\omega}$, $\xi_{JB}$, $\xi_{J\omega}$, $\xi_{SB}$, $\xi_{S\omega}$ depend  at most  three underlying parameters
 through 
\bea
\xi_{JB}&=& C\mu, \nonumber\\
\xi_{J\omega}&=& C\mu^2 +X_B T^2,\nonumber\\ 
\xi_{SB} &=& X_B T,\nonumber\\
\xi_{S\omega}&=& 2\mu T X_B + X_\omega T^2,\nonumber\\
\xi_{TB}&=& \frac 12 \left(C\mu^2 + X_B T^2\right),\nonumber\\
\xi_{T\omega} &=& \frac 23 \left(C\mu^3+3 X_B \mu T^2 +X_\omega T^3\right).
\label{EQ:xi-parameters}
\eea
Here $T$ is the temperature and $\mu$ the chemical potential associated with the ${\rm U}(1)$ particle-number current $J^\mu_N$. 
For the ideal Weyl gas the three parameters $C$, $X_B$, and $X_\omega$ take the values  
\be
C=\frac{1}{4\pi^2},\quad X_B= \frac 1{12},\quad 
X_\omega=0.
\ee
It is clear from the derivation in \cite{no-drag} that  $C$ is the coefficient  of the term (\ref{EQ:anomalyEB})  in  the chiral  anomaly  (\ref{EQ:anom-number}).
It was  conjectured in \cite{landsteiner} that the parameter  $X_B$ is the coefficient  appearing before the Pontryagin density in the same equation. This conjecture was originally motivated by  the simple observation  that both $X_B$ and the Pontryagin coefficient  depend on the same  physical data (spin, chirality, but not charge), but it has  gained support from consideration of global anomalies \cite{golkar,dutta2} and from calculations using  AdS-CFT formalism \cite{landsteinerAdS}. It is not, however, straightforward to confirm the conjecture by extending   the flat-space considerations in \cite{no-drag} to curved space.

For  an ideal gas of  right-handed  Weyl fermions at rest  in flat space (so  $u^\mu=(1,0,0,0)$) the term with $\xi_{TB}$ in (\ref{EQ:energy-anomaly}) leads to an anomaly-induced magneto-thermal energy flux  
\be
{\bf J}_\epsilon= {\bf B} \left(\frac{\mu^2}{8\pi^2} +\frac 1{24} T^2\right),
\label{EQ:magneto-thermal-current}
\ee
which  plays a central role in  \cite{gooth}.

A similar gas in thermal equilibrium in  a frame rotating at angular velocity ${\bf \Omega}$ (so that  $u^\mu=(1,0,0,0)$ on the rotation axis)  acquires from the $\xi_{J\omega}$ term in (\ref{EQ:number-anomaly}) a  chiral vortical effect  (CVE) number current that (again on the rotation axis)  is  given by \cite{vilenkin}
\be
{\bf J}_N = {\bm \Omega}\left( \frac{\mu^2}{4\pi^2}+\frac{ |\Omega|^2}{48\pi^2}+\frac 1{12} T^2\right).
\label{EQ:chiral-vortical-current}
\ee

We do  not need the gravitational anomaly to  understand the origin of the magneto-thermal current in (\ref{EQ:magneto-thermal-current}).
It is well-known  that  solving the for the eigenvalues of the Weyl Hamiltonian in the presence of a magnetic field ${\bf B}$ yields   a set of energy levels
\be
\epsilon_{l}(k)=\pm \sqrt{2|B|l+ k^2},
\ee
where $k$ is the component of momentum parallel to ${\bf B}$. The levels have degeneracy $|B|/2\pi$ per unit area in a plane transverse to ${\bf B}$, 
and all levels are  are gapped except for  $l=0$. The $l=0$  level  is special in that only one sign  of  the square-root is allowed and  we effectively have an array  of gapless one-dimensional chiral fermions with 
\be
\epsilon(k) = +k.
\ee
Each  one-dimensional  chiral fermion contributes a  current of 
\be
J_\epsilon= \int_{-\infty}^{\infty} \frac {d\epsilon}{2\pi} \epsilon \left\{\frac{1}{1+e^{\beta(\epsilon-\mu)}}-\theta(-\epsilon)\right\}=2\pi\left(\frac{\mu^2}{8\pi^2} + \frac{1}{24} T^2\right),
\label{EQ:fermi1}
\ee
where $\beta=1/T$, and the  $-\theta(-\epsilon)$  counter-term effects  a normal-ordering vacuum subtraction of the contribution of  Dirac/Fermi sea, ensuring that   there is no current when $\mu=T=0$. 
Combining (\ref{EQ:fermi1}) with the $|B|/2\pi$ areal degeneracy  leads immediately to (\ref{EQ:magneto-thermal-current}).

The free fermion computation of the CVE current  is rather lengthier  \cite{vilenkin} but it also reduces to a Sommerfeld integral, in this case 
\bea
J_N&=&  \frac 1{4 \pi^2} \int_{-\infty}^\infty   \epsilon^2 \, d\epsilon \left(\frac{1}{1+e^{\beta(\epsilon - (\mu+\Omega/2))}}-\frac{1}{1+e^{\beta(\epsilon - (\mu-\Omega/2))}}\right),\nonumber\\
 &=&\frac{\mu^2 \Omega}{4\pi^2}+\frac {\Omega^3}{48 \pi^2} +\frac 1{12}\Omega T^2.\nonumber
 \label{EQ:chiral-vortical-sommerfeld}
\eea

Although we do not {\it need\/}  the mixed anomalies  to obtain these currents, we {\it can\/} use them  to do so, and in doing so gain insight in the physical origin of the anomalies.   We will devote the next section to  the anomaly derivation of (\ref{EQ:magneto-thermal-current}) and (\ref{EQ:chiral-vortical-current}). We will see that a number of deep ideas are combined  in these derivations.

\section{Currents from anomalies}
\label{SEC:Anomaly-derivations}

In this section we will construct spacetimes in which  the thermal contributions to the anomaly-induced currents  arise from the gravitational  source terms in the associated anomalous conservations laws.

\subsection{magneto-thermal current}
\label{SEC:hawking-anomaly}

 To derive  the thermal part of    (\ref{EQ:magneto-thermal-current})  from the anomaly we will  take  for granted the   $4\to 2$  dimensional reduction provided by the magnetic field, and  consider the current as that of our array of 1+1 dimensional right-going fermions. We imagine a {\it gedanken\/} experiment in which we  heat   each  right-going Fermi field  to  temperature  $T$ by terminating its  space-time on the left by  a 1+1 dimensional  black hole whose  Hawking temperature is $T$. The $T^2$ contribution  to   (\ref{EQ:magneto-thermal-current})  is then the  Fermi field's contribution  to the outgoing Hawking radiation.  To relate this interpretation  to the anomaly  we   review how  \cite{wilczek1,wilczek2,banerjee}  Hawking radiation arises  from   1+1 dimensional version of the energy-momentum anomaly
\be
\nabla_\mu T^{\mu\nu}= -\frac{c}{96\pi } \frac{\epsilon^{\nu\sigma}}{\sqrt{|g|}}\partial_\sigma R,
\label{EQ:2d-grav-anomaly}
\ee
to which  (\ref{EQ:anom-em}) reduces in a uniform ${\bf B}$ field.
Here $\epsilon^{01}=1$, $R=R{^{\alpha\beta}}_{\alpha\beta}=2{R^{01}}_{01}=2{R^{tr}}_{tr}$ is the Ricci scalar, and  $c=c_R-c_L$ is the difference between  the conformal central charges of the right-going and left going massless  fields. As our  magnetic field leaves us with only   right-going  fields, we have $c=1$. 

As the black hole  is  an  externally imposed background space-time, we do  not need its metric   to  satisfy the Einstein equations and a   suitable metric  is   
\be
ds^2 = -f(r)dt^2 + \frac 1{f(r)}dr^2,
\ee
where  all that is required of $f(r)$ is that it  tends to unity  at large $r$ and  vanishes linearly as $r$ approaches the event horizon $r=r_H$. 
In  this metric the Ricci scalar is given by 
\be
R= -f''.
\ee
A  covariant  energy-momentum conservation   equation does not, on its own, lead to conserved  energy and momentum. For that  we need a space-time symmetry {\it i.e.\/}\   a  Killing-vector field $\eta^\mu$ which   obeys   the isometry  condition
\be
\nabla_\mu \eta_\nu+\nabla_\nu \eta_\mu=0.
\ee
Combining the isometry equation with (\ref{EQ:2d-grav-anomaly}) then  gives us  
\be
\nabla_\mu (T^{\mu\nu}\eta_\nu) =  -\frac{c}{96\pi}\frac{\epsilon^{\nu\sigma}\eta_\nu}{ \sqrt{|g|}}\partial_\sigma R,
\ee
in which  
\be
\sqrt{|g|} \nabla_\mu (T^{\mu\nu}\eta_\nu) = \partial_\mu (\sqrt{|g|} T^{\mu\nu}\eta_\nu)
\ee
involves a conventional total divergence.

Our  Schwarzschild-metric posses a Killing vector $\eta=\partial_t$ whose covariant components are $(\eta_t, \eta_r)  =(-f(r),0)$. From this we find that  
\be
\nabla_\mu T^{\mu\nu } \eta_\nu  = (\partial_r \sqrt {|g|} {T^r}_t )/\sqrt{-g}
\ee
 We then have 
\be
\frac{\partial}{\partial r}( \sqrt {|g|} {T^r}_t)=-\frac c {96\pi} f\partial_ r f'' = -\frac{c}{96\pi} \frac{\partial}{\partial r} \left(ff''-\frac 12(f')^2\right),
\ee
and integrating from $r_H$ to $r=\infty$ gives 
\be
\left. \sqrt{|g|} {T^r}_t \right|^\infty_{r_H} = -\left.\frac{c}{96\pi} \left(ff''-\frac 12(f')^2\right)\right|^\infty_{r_H}.
\label{EQ:integrated-anomaly}
\ee
According to   \cite{banerjee},   the appropriate boundary condition is that  ${T^r}_t $ be   zero at the horizon.  The RHS of (\ref{EQ:integrated-anomaly}) by contrast is zero at infinity, and contributes $(c/96\pi) (f')^2/2$  at  the horizon.
As $\sqrt{|g|} \to 1$ at large $r$, we see that an  energy current of magnitude 
\be
T^{rt} (z\to \infty)=- {T^r}_t(z\to \infty) =\frac{c\kappa^2 }{48\pi}, \quad \kappa= f'(r_H)/2,
\ee
has been built up  by the anomaly  as we move away from the horizon. 
The quantity  $\kappa$  is the {\it surface gravity\/} of the black hole. 

\begin{figure}
\includegraphics[height=5.0cm]{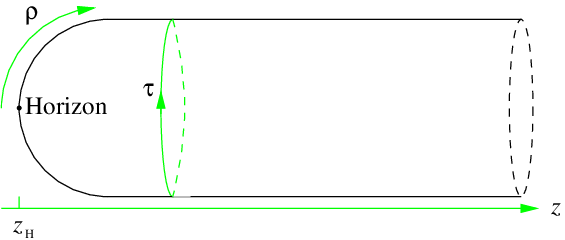}
\caption{\sl The Euclidean, imaginary time,  section  of the 2-dimensional black hole is asymptotically a cylinder of circumference $2\pi/\kappa$. The horizon is a single point at $\rho=0$, or  equivalently $z=z_H$.} 
\end{figure}

To  complete the derivation of (\ref{EQ:magneto-thermal-current}) we  recall  the argument \cite{hawking-hartle,gibbons-perry}  that the geometry of the  Euclidean section of our black-hole metric shows that the Hawking temperature is given by  $T_H=\kappa/2\pi$.  We begin by setting  
 $t=-i\tau$ and   see that in  imaginary-time  our Schwarzschild space metric becomes 
\be
d\sigma^2 = f(r) d\tau^2 +\frac 1 {f(r)} dr^2.
\label{EQ:euclidean-schwarzschild}
\ee
If we introduce a new radial co-ordinate 
\be
\rho = \int_{r_H}^r \frac{dr'}{\sqrt{f(r')}}\approx \frac{2}{\sqrt{f'(r_H)}}\sqrt{r-r_H},
\ee
where the approximation holds for $r$  just above $r_H$.
Then, in this same region,
\bea
d\sigma^2 &=&f(r)d\tau^2+\frac 1{f(r)}dr^2\nonumber\\
&=& f(r) d\tau^2 +d\rho^2,\nonumber\\
&\approx& f'(r_H) (r-r_H)d\tau^2+d\rho^2\nonumber\\
&=&\kappa^2  \rho^2d\tau^2+d\rho ^2.
\eea
Comparison with the metric of plane polar coordinates now shows that if  there is to be no conical singularity at $r_H$  we must identify
$
\kappa \tau
$
with the polar angle $\theta$. 
Thus the smooth euclidean manifold described by (\ref{EQ:euclidean-schwarzschild}) looks like the skin  of a salami sausage in which  the circumferential coordinate  $\theta$ is identified with $\theta+2\pi$, or equivalently  $\tau$ is  identified with $\tau+\beta$ where  $\beta= 2\pi/\kappa$.  Green functions $G(r,t)$ in Minkowski signature spacetime will be periodic in imaginary time with period $\beta$ and are therefore \cite{hawking-hartle,gibbons-perry}  thermal Green functions with temperature $T_H=\beta^{-1}=\kappa/2\pi$ (or $k_{\rm B}T_H = \hbar \kappa/2\pi c$ if we include dimensionful constants). 

This derivation seems quite straightforward, but there are two  subtleties that need to be discussed. Firstly,  anomalies are usually presented as being of two types: {\it consistent\/}  and {\it covariant\/} \cite{bardeen-zumino}. Following \cite{banerjee} we  have exclusively used the covariant form of the anomaly. Secondly, it is well known that Hawking radiation is observer-dependent.  These two issues are not unrelated. To illuminate  this point we will repeat  the Hawking radiation calculation using the two-dimensional version of Kruskal-Szekeres coordinates.

We begin by defining  a {\it tortoise coordinate\/} $r_*$ by solving 
\be 
\frac{dr_*}{dr} = \frac 1 {f(r)},
\ee
and taking as boundary condition that   $r_*$ coincides with $r$ at large positive distance.  
In    $(t,r_*)$ coordinates the metric becomes 
\be
ds^2 = e^\phi (-dt^2+dr_*^2)
\ee
where 
\be
\phi(r_*)= \ln f(r(r_*))
\ee
and the  event horizon lies   at $r_*=-\infty$. 
On setting   $z=r_*+i\tau$ and $\bz = r_*-i\tau$  the euclidean version of this metric  takes the  isothermal (conformal)  form    
\be
d\sigma^2 = e^{\phi} d\bz\,dz.
\ee
In  $\bz, z$   coordinates  the only non-vanishing Christoffel symbols are 
\be
\Gamma_{zz}^z= \partial_z \phi, \quad \Gamma_{\bz\bz}^\bz= \partial_\bz \phi,
\ee
and the Ricci scalar is 
\be
R= -4e^{-\phi} \partial^2_{z\bz} \phi.
\ee

In  two-dimensional conformal field theory we are used to defining  energy-momentum operators $\hat T(z)$ and $\hat {\bar T}(\bz)$ where, for a free $c=1$ boson field  $\hat \varphi(z,\bz)= \hat  \varphi(z)+\hat \varphi(\bz)$ for example,   we have 
\bea
\hat T(z) &=&: \partial_z \hat\varphi(z) \partial_z\hat \varphi(z) :\nonumber\\
&=& \lim_{\delta  \to 0}\left( \partial_z \hat \varphi(z+\delta/2) \partial_z \hat \varphi(z-\delta/2)+ \frac{1}{4\pi \delta^2}\right). 
\label{EQ:Tz}
\eea 
(Note that conformal field theory  papers often  define $\hat T(z)$ to be $-2\pi$ times (\ref{EQ:Tz}) so as to simplify the operator product expansion.)
The operator $\hat T(z)$  has been constructed to be explicitly holomorphic in $z$,  but at a price of  tying its   definition  to the $z$, $\bz$ coordinate system --- both in  the  normal ordered  expression in  the first line  and by the explicit counterterm in the second line. It is not surprising, therefore, that under  a  change of co-ordinates  the operator $\hat T(z)$ does not transform as a tensor but instead acquires a inhomogeneous Schwarzian-derivative c-number part \cite{difrancesco}. If we want a genuine  energy-momentum   {\it tensor\/} we must define 
\bea
\hat T_{zz}&=& \hat T(z)+      \frac {c}{24 \pi}  \left(\partial^2_{zz} \phi- \frac 12 (\partial_z \phi)^2\right),\nonumber\\
\hat T_{\bz\bz}&=& \hat {\bar T}(\bz)+      \frac {c}{24 \pi}  \left(\partial^2_{\bz\bz} \phi- \frac 12 (\partial_\bz \phi)^2\right),\nonumber\\
\hat T_{\bz z}&=& -\frac {c}{24  \pi} \partial^2_{z\bz} \phi,
\label{EQ:fullT}
\eea 
in which the c-number Schwarzians  in  the operator transformation  are  cancelled by   Schwarzians from the transformation of the c-number additions.   

A direct computation, using the holomorphicity and anti-holomorphicity of the operators $\hat T(z)$ and $\hat {\bar T}(\bz)$ together with  the formul{\ae}\  for the Christoffel symbols, shows that 
\be
  \nabla^z \hat T_{zz}+\nabla^\bz \hat T_{\bz z}=0.
  \ee
  If, however,  we keep   only the right-going field,  the chiral energy momentum tensor becomes 
  \bea
\hat T_{zz}&=& \hat T(z)+      \frac {c}{24 \pi}  \left(\partial^2_{zz} \phi- \frac 12 (\partial_z \phi)^2\right),\nonumber\\
\hat T_{\bz\bz}&=& 0,\nonumber\\
\hat T_{\bz z}&=& -\frac {c}{48  \pi} \partial^2_{z\bz} \phi.
\label{EQ:chiralT}
\eea 
 A  similar  computation  shows that  the chiral tensor  obeys 
  \bea
  \nabla^z \hat T_{zz}+\nabla^\bz \hat T_{\bz z}=- \frac{c}{96\pi}\partial_z R,\nonumber \\
  \nabla^z\hat T_{z\bz } +\nabla^\bz \hat T_{\bz \bz}= +\frac{c}{96\pi}\partial_\bz R,
 \eea
 where the second term on the left hand side of  the second equation is  identically  zero.
 In our  $z$, $\bz$ co-ordinates system we have  $\sqrt{g} = \sqrt{-g_{\bz z}g_{z\bz}}=- i e^\phi/2$ (perhaps more clearly, we can express this  as   $e^\phi dt \wedge dr_*= (e^\phi/2i) dz \wedge d \bz$), and  we  can write these last two equations  in a covariant manner as   
\bea
  \nabla^z \hat T_{zz}+\nabla^\bz \hat T_{\bz z}=i \frac{c}{96\pi}\sqrt{g}\,\epsilon_{z\bz}  \partial^\bz R, \nonumber\\
  \nabla^z\hat T_{z\bz } +\nabla^\bz \hat T_{\bz \bz}= i\frac{c}{96\pi}\sqrt{g}\,\epsilon_{\bz z} \partial^z R.
 \eea
 In general euclidean co-ordinates we therefore have \cite{fulling} 
 \be
 \nabla^\mu \hat T_{\mu\nu} = i\frac{c}{96 \pi} \sqrt{g}\, {\epsilon_{\nu \sigma}} \partial^\sigma R.
 \label{EQ:2dg_euclidanomaly}
 \ee
 The  factor ``$i$''  appears in (\ref{EQ:2dg_anomaly}) because it is only the {\it imaginary\/} part of the Euclidean effective action that can be anomalous \cite{alvarez-gaume-witten,alvarez-gaume-ginsparg}.  It is absent when we write the equation in Minkowksi signature space-time  where it becomes
 \be
 \nabla_\mu \hat T^{\mu\nu} = -\frac{c}{96 \pi} \frac{1}{\sqrt{|g|} }{\epsilon^{\nu \sigma}} \partial_\sigma R.
 \label{EQ:2dg_anomaly}
 \ee
 Because  we have used a covariantly-transforming energy momentum tensor, we find the {\it covariant\/} form of the anomaly.  
In this calculation we also see that  the anomaly arises solely from the c-number terms. 

Now define  Euclidean Kruskal coordinates $U$, $V$ by setting
\bea
Z=U+iV&=& \exp\{ \kappa(r_*+i\tau)\}= \exp\{\kappa z\},\nonumber\\
\bar Z=U-iV &=&\exp\{ \kappa(r_*-i\tau)\}=\exp\{\kappa \bz\},
\eea
so that 
\be
|Z|^2=U^2+V^2= \exp\{2\kappa r_*\}.
\ee
In terms of these coordinates  we have 
\be
d\sigma^2= f(r) \kappa^{-2} e^{-2\kappa r_*}(dU^2+dV^2).
\ee
With $\kappa$ being the surface gravity, this goes  to the non-singular metric
\be
d\sigma^2={\rm const.} (dU^2+dV^2)
\ee
(where  the constant is determined by  the exact form of $f(r)$) near the horizon point at $U^2+V^2=0$) and to
\be
d\sigma^2= \kappa^{-2} (U^2+V^2)^{-1} (dU^2+dV^2)
\ee
at large distance.
This last expression is the metric of a cylinder of circumference $2\pi/\kappa$,  confirming the time periodicity.

For a general conformal ``salami sausage''  metric we need $ds^2= e^\phi dZd\bar Z$ with $\phi=0$ at $|Z|=0$, and $\phi\approx -2\ln \kappa|Z|$ at large $|Z|$ where the circumference of the sausage becomes constant. The coefficient ``$-2$'' is required by the Gauss-Bonnet theorem as the end-cap is  topologically a hemisphere. At short distance 
the sausage looks like a spherical cap, and we have 
\be
e^\phi= 1-\frac 1{4} \bz z R+O(|z|^4), \quad \phi= -\frac 1{4} \bz z R+O(|z|^4),
\ee
where $R$ is the Ricci scalar (twice the Gaussian curvature) at the horizon.

As  Kruskal $Z$, $\bar Z$ coordinates are  again  isothermal, the chiral energy momentum tensor is  of the form
\be
\hat T_{ZZ}= \hat T(Z)+\frac c{24\pi} \left(\partial^2_{ZZ} \phi-\frac 12 (\partial_Z \phi)^2\right),
\ee
where $\hat T(Z)$ is the normal-ordered operator part that transforms inhomogeneously under conformal maps. The second term is the c-number counterterm  whose transformation  cancels that of $\hat T(Z)$ so as to make $\hat T_{ZZ}$  transform as a  tensor.

At  short distance   
the c-number part in $\hat T_{ZZ}$ vanishes ---  in fact  it vanishes identically on a sphere with conformal coordinates.  Consequently, as    $\hat T_{ZZ}$ is zero at the horizon,  the  expectation value of  $\hat T(Z)$  is  zero there, and hence   everywhere.   
At large distance, however,  we will have
\be
\phi(Z,\bar Z)\sim -\ln \kappa Z-\ln\kappa \bar Z
\ee
and so the $c$-number part gives us 
\bea
T_{ZZ}&\sim& \frac{c}{24\pi}\left(\frac 1{Z^2}-\frac 12 \frac 1{Z^2}\right)\nonumber\\
&=& \frac{c}{48\pi}\frac{1}{Z^2}.
\eea
Now let us shift to the tortoise light-cone coordinates $z=r_*+i\tau $, $\bz=r_*-i\tau$. Then  
\bea
\hat T_{zz} &=& \left(\frac{\partial Z}{\partial z}\right)^2 \hat T_{ZZ}\nonumber\\
&=& \kappa^2 Z^2 \hat T_{ZZ}\nonumber\\
&\to& \frac {c\kappa^2}{48\pi}, \quad \hbox{as $r_*\to \infty$}.
\eea
In the asymptotic Minkowski space $r_*=r$, and with the speed of light equal to one and $\pm$ denoting $r\pm t$ components, we have 
\bea
\hat T_{++} &=& \frac 14 (\hat T_{rr}+\hat T_{tt}-2\hat T_{rt}),\nonumber\\
\hat T_{--}&=& \frac 14 (\hat T_{rr}+\hat T_{tt}+2\hat T_{rt}),\nonumber\\
\hat T_{+-}&=& \frac 14(\hat T_{rr}-\hat T_{tt}),
\eea
with $\hat T_{+-}=\hat T_{--}=0$. Consequently  $\hat T_{++}=\hat T_{tt}=\hat T_{rt}$ and the Kruskal coordinate energy density  and   flux coincide with those from the  Schwarzschild coordinate calculation.
The break-up between operator and c-number is different  however.
The $c$-number part in $\hat T_{zz}$   vanishes at large $r$   so the large-distance contribution to the Schwarzschild energy flux  comes entirely from the expectation-value  of the  operator $\hat T(z)$. In other words, the Schwarzschild observer sees the  asymptotic energy  being carried by actual particles.  In  Kruskal coordinates the $\hat T_{ZZ}$  operator part has vanishing expectation value everywhere and the asymptotic energy flux comes entirely from the c-number term. Thus $\hat T(Z)$  and $\hat T(z)$ record  very different particle content and the Schwarzschild $r$, $t$ coordinate  observer's zero-particle state  is  not the same as the Kruskal observer's zero-particle state.  

The physical interpretation  should now be clear:  Both the Schwarzschild time  and the Kruskal time coordinate lines correspond to  the flows of   time-like Killing vectors. In each coordinate system the field's  mode expansions have   well-defined yet different  positive-frequency modes whose coefficients are  annihilation operators. The operators $\hat T_{ZZ}$ and $\hat T_{zz}$ are normal ordered so that the annihilators are all to the right.  It is the   positive frequencies  that can excite a  detector from its ground state, and  the normal-ordered operators keep track of what a detector at  fixed spatial coordinate  would record   in each coordinate system. Close to the horizon a detector at fixed Schwarzschild  coordinate $r$ sees a high-temperature thermal distribution of outgoing particles. However,   at the horizon, their contribution to the expectation value of $\hat T_{z}$ is exactly  cancelled by the c-number term.  As we move away from the horizon this  c-number term decreases  and allows the total covariantly-defined energy current to grow to its asymptotic value. On the other hand the  Kruskal obsrever   never sees any particles, and all their     energy  flux comes from the c-number contribution which  grows from zero at the horizon to  the {\it same\/}  asymptotic value as the Schwarzschild flux. Presumably the source term in the Einstein equations will  be the expectation value of a   covariantly-defined   energy-momentum  tensor.  Therefore it is independent of the observer's motion ---  but as  we are not   investigating   the back reaction of the emitted radiation on the  geometry, this is not our present concern.    

\subsection{Chiral Vortical current}
\label{SEC:chiral-vortical-anomaly}
  
We now seek an analogous derivation of  (\ref{EQ:chiral-vortical-current}) from the Pontryagin term in (\ref{EQ:anom-number}).
To do this we  need to modify  our  toy black hole  to acquire  a   non-zero Pontryagin density. 
 A suggestion of how to proceed comes from the Kerr metric of a rotating black hole.  In  Boyer-Lindquist coordinates $(t,r,\phi, \theta)$, and with  $\cos\theta=\chi$, this metric is     
 \bea
ds^2&=&-\left(1-\frac{2mr}{r^2+a^2\chi^2}\right)dt^2+ \left(\frac{r^2+a^2\chi^2}{r^2+a^2-2mr}\right) dr^2 -\frac{ 4 a m r(1-\chi^2)}{r^2+a^2\chi^2} dt d\phi\nonumber\\
&&(1-\chi^2)\left( a^2+r^2 +\frac{2a^2 mr(1-\chi^2)}{ r^2+a^2 \chi^2}\right) d\phi^2 + \frac{r^2+a^2 \chi^2}{1-\chi^2}d\chi^2,
\eea
where $m$ is the mass and $J=ma$  the angular momentum of the black hole.

This  metric has two special horizon surfaces at the roots
\be
r_{\pm}= m\pm\sqrt{m^2-a^2}
\ee
of $r^2+a^2-2mr=0$. The outer horizon $r=r_+$ is the causal event horizon on which trapped photons are forced to orbit  at  fixed $r, \theta$ and    angular velocity     
\be
\Omega_+ \stackrel{\rm def}{=}\frac{d\phi}{dt} = \frac{a}{r_{+}^2}.
\ee
Both $\Omega_+$  and the surface gravity 
\be
\kappa_+ = \frac{1}{4m}- m\Omega_+^2 
\ee
are  constant over the horizon.  As in the Schwarzschild case, the  absence of a conical singularity in the Euclidean section requires $\tau \sim \tau+  \beta_H$ where  \cite{euclidean-kerr} 
 \be
 \beta_H= \frac{1}{T_H} = \frac{2\pi}{\kappa_+}.
 \ee
 The Kerr black hole is therefore both rotating and hot.
 
What is important for us is that the  numerical coefficient 
\be 
\frac 1 4 \epsilon^{\lambda\mu\rho\sigma} {R^a}_{b\lambda\mu} {R^b}_{a\rho\sigma} 
= - \frac{48 a m^2 r \chi(r^2- 3a^2 \chi^2)(3r^2- a^2 \chi^2)}{(r^2+a^2 \chi^2)^5}
\ee   
of $dt\wedge dr\wedge d\phi\wedge d\chi$ in the Pontryagin-density 4-form $\tr\{R^2\}$ is non-zero.  For small $a/m$ the coefficient is largest 
near the poles of rotation at $\chi=\pm 1$. 

The Kerr  metric can be conveniently written \cite{euclidean-kerr} in terms of the functions 
\bea
\Delta &=& r^2+a^2-2mr,\\
\rho^2 &=& r^2+a^2 \cos^2\theta,
\eea
and two 
mutually orthogonal one forms 
\bea
\omega&=& \frac{r^2+a^2}{\rho ^2}(dt-a\sin^2\theta d\phi),\\
\tilde\omega&=& \frac{r^2+a^2}{\rho ^2}\left(d\phi-\frac{a}{r^2+a^2}  dt\right),
\eea
as 
\be
 ds^2=- \frac{\Delta \rho^2}{(r^2+a^2)^2}\omega^2 + \frac{\rho^2}{\Delta} dr^2 +\rho^2 d\theta^2 +\rho^2\sin^2\theta \tilde\omega^2.
 \ee
 Motivated by this rewriting, we consider a $3+1$ space with metric  
\be
ds^2 = -f(z) \frac{\left(dt-\Omega\, r^2 d\phi \right)^2}{(1-\Omega^2 r^2)} +\frac 1{f(z)}dz^2 + dr^2+\frac{ r^2\left(d\phi -\Omega \,dt \right)^2}{(1-\Omega^2 r^2)}.
\label{EQ:our-metric}
\ee
 The   metric (\ref{EQ:our-metric}) has been constructed so that at large $z$, where $f(z)=1$,   it reduces to 
 \be
 ds^2\to -dt^2 +dz^2 + dr^2 +r^2 d\phi^2
 \ee
where   $r$,  $z$, and $\phi$ and $z$  are the   cylindrical-coordinate  radial, axial, and azimuthal   coordinates  of an asymptoticly flat   spacetime. Thus,  in contrast to previous usage, $z$  is a real coordinate that provides  a measure of the distance from the horizon at $f(z_H)=0$, and $r$  is a measure of  distance from the rotation axis.  Replacing the complicated Kerr-metric  coefficients by the function $f(z)$  allows us to ensure that  the space-time curvature is   concentrated near the horizon, which now appears to be planar and rotating at angular velocity $\Omega$.  We anticipate  that outgoing  Fermi fields  in this space  will be in asymptotic thermal equilibrium   at temperature $T_H= f'(z_H)/4\pi$ in a frame rotating about the $z$-axis with the horizon  angular velocity $\Omega$. They should therefore acquire  a  CVE current as they move though the curved and twisted near-horizon geometry.     

The numerical coefficient of the Pontryagin density in this space-time is
\bea
\frac 14 \epsilon^{\mu\nu\rho\sigma} {R^\alpha}_{\beta\mu\nu} {R^{\beta}}_{\alpha\rho\sigma}&=& \frac{2r \Omega f'(z)(8\Omega^2(1-f(z))+(1-\Omega^2r^2)^2f''(z))}{(1-\Omega^2 r^2)^3}
\nonumber\\
&\sim & 2r \Omega  f'(z)f''(z) \nonumber\\
&=& \frac{\partial}{\partial z} ( \Omega r [f'(z)]^2).
\eea
In the last two lines we have kept  only the leading term in $\Omega$. The error is  $O[\Omega^3]$. 

When we divide by $\sqrt{|g|}=r$ to get the Pontryagin-density scalar we see that we have  created in the region abutting  the  horizon  an $r$-independent   source term   for the axial current of our anomalous relativistic fluid. We assume that  this planar source drives a current only in the $z$ direction. In that case, we  
find that 
to  leading order in $\Omega$  the anomalous conservation law  (\ref{EQ:anom-number}) becomes 
\be
 \frac{\partial}{\partial z}( \sqrt{|g|} J^z_N)= -\frac{ \Omega} {192  \pi^2 } \frac{\partial}{\partial z}( \sqrt{|g|}  [f'(z)]^2).
\ee
With boundary condition $J^z(z_H)=0$, we can again integrate up with respect to $z$ to find 
\be
 J_N^z(z\to \infty)= \frac{1}{12} \Omega T^2_H.
 \label{EQ:CVE-from-anomaly}
\ee
This  is the expected  thermal contribution to the CVE current (\ref{EQ:chiral-vortical-current}). 

If we retain terms of order $\Omega^3$, we do get a contribution to the on-axis current similar to that in (\ref{EQ:chiral-vortical-current}), but with with coefficient $1/24\pi^2$ rather than $1/48\pi^2$.  Trying slightly modified metrics indicates that this correction to the  current is sensitive to  how the metric varies away from the axis of rotation.  
For example, omitting the $(1-\Omega^2 r^2)$ factors in the denominators in (\ref{EQ:our-metric}) does not alter the on-axis asymptotic metric, and does not affect  the coefficient of the $T^2$ term in the current. It does, however,  lead to the  coefficient of
 $\Omega^3$ becoming  zero. Perhaps we should not be surprised by this as the notion of a rigidly rotating coordinate system such as that  used by \cite{vilenkin} is bound to be problematic away  from the rotation axis.

Note that our CVE  current (\ref{EQ:CVE-from-anomaly}) is not, as suggested in \cite{flachi},  simply proportional to   the    Chern-Simons current associated with the Pontryagin class.  The latter current  
\be
J_{\rm CS}^\lambda = \frac 12 \frac{\epsilon^{\lambda\mu\rho\sigma}}{\sqrt{|g|}}\left( {\Gamma^\alpha}_{\beta\mu} \partial_\rho {\Gamma^\beta}_{\alpha \sigma}+ \frac 23 {\Gamma^\alpha}_{\beta\mu} 
{\Gamma^\beta}_{\gamma\rho}{\Gamma^\gamma}_{\alpha\sigma}\right)  
\ee
 has (to leading order in $\Omega$) {\it two\/} non-zero components 
\be 
J^z_{\rm CS}= \Omega  [f'(z)]^2, \quad J^r_{\rm CS}= 2 \Omega f'(z)/r.
\ee
It does satisfy 
\be
\partial_\lambda \sqrt{|g|} J^\lambda_{\rm CS} = \frac 1 4 \epsilon^{\lambda\mu\rho\sigma} {R^\alpha}_{\beta\lambda\mu} {R^\beta}_{\alpha\rho\sigma}, 
\ee
but obeys different boundary conditions in that $J^\mu_{\rm CS}(z)$  vanishes at $z=\infty$  rather than at the horizon.  Our derivation  in this section  was, however, motivated  by the discussion    in \cite{flachi}. 

\section{Sommerfeld integrals and Anomalies}  
\label{SEC:generating}

In the introduction we  made the  claim that the Fermi-distribution moment integrals that appear  in the higher-order terms of the  Sommerfeld expansion somehow know about  anomalies.  In this section we try to explain how this knowledge comes about by combining  the ideas in \cite{ideal} with the geometry behind our  {\it gedanken\/} trick of using the Hawking effect as our heat source.   

The first few such moment integrals are 
\bea
\int_{-\infty}^\infty \frac{d\epsilon}{2\pi}\left\{ \frac{1}{1+e^{\beta(\epsilon-\mu)}}-\theta(-\epsilon)\right\} &=&\left(\frac{\mu}{2\pi}\right),\nonumber\\
\int_{-\infty}^\infty \frac{d\epsilon}{2\pi}\left(\frac{\epsilon}{2\pi}\right)\left\{ \frac{1}{1+e^{\beta(\epsilon-\mu)}}-\theta(-\epsilon)\right\} &=&\frac{1}{2!}\left(\frac{\mu}{2\pi}\right)^2+\frac{T^2}{4!},\nonumber\\
\int_{-\infty}^\infty \frac{d\epsilon}{2\pi}\frac 1{2!}\left(\frac{\epsilon}{2\pi}\right)^2\left\{ \frac{1}{1+e^{\beta(\epsilon-\mu)}}-\theta(-\epsilon)\right\} &=&\frac{1}{3!}\left(\frac{\mu}{2\pi}\right)^3+\left(\frac{\mu}{2\pi}\right)\frac{T^2}{4!},\nonumber\\
\int_{-\infty}^\infty \frac{d\epsilon}{2\pi}\frac1{3!}\left(\frac{\epsilon}{2\pi}\right)^3\left\{ \frac{1}{1+e^{\beta(\epsilon-\mu)}}-\theta(-\epsilon)\right\} &=&\frac{1}{4!}\left(\frac{\mu}{2\pi}\right)^4+\frac{1}{2!}\left(\frac{\mu}{2\pi}\right)^2\frac{T^2}{4!}+\frac 78\frac{T^4}{6!},\nonumber\\
\int_{-\infty}^\infty \frac{d\epsilon}{2\pi}\frac 1{4!} \left(\frac{\epsilon}{2\pi}\right)^4\left\{ \frac{1}{1+e^{\beta(\epsilon-\mu)}}-\theta(-\epsilon)\right\} &=&\frac{1}{5!}\left(\frac{\mu}{2\pi}\right)^5+\frac{1}{3!}\left(\frac{\mu}{2\pi}\right)^3\frac{T^2}{4!}+\left(\frac{\mu}{2\pi}\right) \frac 78\frac{T^4}{6!},\nonumber\\
\int_{-\infty}^\infty \frac{d\epsilon}{2\pi}\frac 1{5!}\left(\frac{\epsilon}{2\pi}\right)^5\left\{ \frac{1}{1+e^{\beta(\epsilon-\mu)}}-\theta(-\epsilon)\right\}&=&\frac{1}{6!}\left(\frac{\mu}{2\pi}\right)^6+\frac{1}{4!}\left(\frac{\mu}{2\pi}\right)^4\frac{T^2}{4!}\nonumber\\
&&\quad +\frac{1}{2!} \left(\frac{\mu}{2\pi}\right)^2 \frac 78\frac{T^4}{6!}+ \frac{31}{24} \frac{T^6}{8!}.
\label{EQ:a_few_Sommerfeld}
\eea
These   are all polynomials in the temperature and the chemical potential. 
It is essential for the simplicity of these results that the $\epsilon$ integral runs from $-\infty$ to $+\infty$. If we had kept only the positive energy part of the integrals we would have  
instead 
\be
\int_0^{\infty}\frac{d\epsilon}{2\pi}\frac 1{k!}\left( \frac{\epsilon}{2\pi}\right)^k \frac{1}{1+e^{\beta(\epsilon-\mu)}}= - \frac{1}{(2\pi  \beta)^{k+1}}{\rm Li}_{k+1}(-e^{\beta\mu}),
\ee
where the polylogarithm function ${\rm Li}_{k}(x)$ is defined by analytic continuation from the series 
\be
{\rm Li}_{k}(x)= \sum_{n=1}^\infty  \frac{x^n}{n^k}, \quad |z|<1.
\ee  
The polynomial form  of the full-range  integral arises from the identity
\be
{\rm Li}_{k}(-e^{\beta\mu})+(-1)^k{\rm Li}_{k}(-e^{-\beta\mu})=-\frac{(2\pi i)^k}{k!}{\rm B}_k\left(\frac 12+\frac{\beta\mu}{2\pi i}\right)
\label{EQ:hurwitz}
\ee
which holds for integer $k$, and where   ${\rm B}_k(x)$ are the Bernoulli polynomials defined by
\be
\frac{t e^{tx}}{e^t-1}= \sum_{n=0}^\infty \frac{t^n}{n!} B_n(x).
\ee
The identity (\ref{EQ:hurwitz})  is a special  case  of a  general  identity for the polylogarithm due to Hurwitz.
A compact generating function 
\be
\int_{-\infty}^{\infty} \frac{d\epsilon}{2\pi} e^{\tau\epsilon/2\pi} \left\{ \frac{1}{1+e^{\beta(\epsilon-\mu)}}-\theta(-\epsilon)\right\}= \frac 1{\tau}\left\{ \frac{(\frac{\tau T}{2})}{\sin(\frac{\tau T}{2})} e^{\tau\mu/2\pi}-1\right\}, \quad 0<\tau T/2\pi< 1,
\label{EQ:generating-function}
\ee
 for the Fermi-distribution moments encapsulates these facts. Expanding both sides  of (\ref{EQ:generating-function})  in powers of $\tau$ and comparing coefficients reveals the equalities  in (\ref{EQ:a_few_Sommerfeld}), and also explains the reason for the  inclusion of the factors of   $1/n! (2\pi)^n$ in the left hand side integrals  of (\ref{EQ:a_few_Sommerfeld}).  The generating function identity 
 (\ref{EQ:generating-function}) is easily established  by  substituting   $x= \exp\{\beta(\epsilon-\mu)\}$ and then using the standard integral  
\be
\int_0^\infty dx \frac{x^{\alpha-1}}{1+x}= \frac{\pi}{\sin\pi \alpha}, \quad 0<\alpha<1.
\ee

The authors of  \cite{ideal} point out that the generating function (\ref{EQ:generating-function}) is strongly   reminiscent of the  general formula 
\be
{\rm Index}[\slashed D]=\int_{\mathcal M} \hat A[R]\, {\rm ch}[F]
\label{EQ:dirac-index}
\ee
for the index of the Dirac operator on a euclidean manifold ${\mathcal M}$. Here 
\be
{\rm ch}[\tau F]= \exp\{\tau F/2\pi i\}= 1+ \tau \{F/2\pi i\}+ \frac{\tau^2} {2} \{F/2\pi i\}^2+\ldots
\ee
is the total  Chern character involving the gauge-field curvature $F= {\textstyle \frac 12} F_{\mu\nu} dx^\mu dx^\nu$, and  
\bea
\hat A[\tau R]&\stackrel{\rm def}{=}& \sqrt{{\rm det}\!\left(\frac{\tau R/4\pi i}{\sinh \tau R /4\pi i}\right)} \nonumber\\
 &=&1+ \frac{\tau^2 }{(4\pi)^2} {2}\tr\{ R^2\} +{\frac {\tau^4 }{(4\pi)^4}}\left[\frac{1}{288} (\tr \{R^2\})^2+ \frac{1}{360}\tr \{R^4\}\right]+\cdots\nonumber\\
&=& 1-  \frac {\tau^2}  {24}{\mathfrak p}_1 + \frac {\tau^4}{5760}(7{\mathfrak p}_1^2-4{\mathfrak p}_2)+\ldots.
\label{EQ:A-roof1}
\eea
is the A-roof genus involving the Riemann curvature matrix-valued  two-form
\be
R_{ij}=\frac 12 R_{ij\mu\nu} dx^\mu dx^\nu.
\ee
In the last line of (\ref{EQ:A-roof1}) the  4N-forms ${\mathfrak p}_n(R)$ are the Pontryagin classes  normalized as   is customary in the mathematics literature. It is tacitly  understood  that  in the product of ${\rm ch}[F]$ and $\hat A(R)$  in (\ref{EQ:dirac-index})  we only retain those terms whose total  form degree matches that of the manifold ${\mathcal M}$.

To derive the equalities in  (\ref{EQ:A-roof1}) and see  the connection with (\ref{EQ:generating-function}) we make use of the algebraic trick that underlies    the {\it splitting principal \/} from the general theory of characteristic classes.  We regard the  curvature two-form of the  $N$-dimensional manifold ${\mathcal M}$  as an $n$-by-$n$  skew-symmetric matrix that can be reduced to the canonical form  
\bea
\frac 1{2\pi}  R_{ij}\equiv  \frac 1 {4\pi}R_{ij\mu\nu}dx^\mu dx^\nu= \left[\matrix{0&-x_1 & & & &\cr x_1 &0 & & & & \cr  & &\ddots &&& \cr &&  &0 &-x_{N/2} \cr &&& x_{N/2}&0}\right]_{ij}.
\eea
Here the  $x_i$ are formal objects (Chern roots)  which  become real  numbers when we evaluate the curvature two-form  at a point on some chosen vectors, and only then perform the canonical-form reduction. In terms of the $x_i$, the A-roof genus and the total Pontryagin class are given  by  
\bea
\hat A[\tau R]&=& \sqrt{\det\left(\frac{\tau R/4\pi i}{\sinh \tau R /4\pi i}\right)}\,= \prod_{i=1}^{n/2} \frac{\tau x_i/2}{\sinh \tau x_i/2},
\label{EQ:A-roof2}
\\
{\mathfrak  p}(\tau R)&=&\quad \det(1-\tau R/2\pi) \qquad = \prod_i (1+\tau^2 x_i^2),
\eea
and 
\be
{\mathfrak  p}(\tau R)=1+\tau^2 {\mathfrak p}_1(R)+ \tau^4 {\mathfrak p}_2(R)+\ldots.
\ee
For the Pontryagin classes the expressions 
\bea
{\mathfrak p}_1(R)&=& \sum_ix_1^2= - \frac 1{(2\pi)^2}\left[\frac 12 \tr \{R^2\}\right],\nonumber\\
{\mathfrak p}_2 (R)&=& \sum_{i<j} x_i^2x_j^2=\frac 1{(2\pi)^4}\left[\frac 18 (\tr \{R^2\})^2- \frac 1 4\tr\{R^4\}\right],  
\eea account for the equality of the last two lines in (\ref{EQ:A-roof1}). A similar expansion of (\ref{EQ:A-roof2}) leads to the equality of the first two lines.

The discussion in  \cite{ideal} combines a general solution   \cite{loga}  to   the  constraints imposed by demanding absence of entropy creation by the anomaly induced currents   with the generating function (\ref{EQ:generating-function}) to   obtain an   effective action for the  ideal Weyl gas  from the anomaly polynomial ${\mathfrak P}[R,F]\stackrel{\rm def}{=} \hat A[R]\, {\rm ch}[F]$.   A key ingrediant  is  the 
 replacement rule \cite{ideal,azeyanagi,banarjee-loga,jensen1,jensen2,jensen3,stone-dwivedi}
\bea
F&\to& \mu\nonumber\\
{\mathfrak p}_1(R) &=& - \frac{1}{8\pi^2} \tr (R^2) \to -T^2 \nonumber\\
{\mathfrak p}_n(R) &=& 0, \quad n>1.
\eea
 
 The replacement rule result is very striking but one is left wondering whether the similarity of the Sommerfeld-integral generating function's factor 
\be
 \frac{{\tau T}/{2}}{\sin({\tau T}/{2})} 
 \ee
 to the anomaly polynomial's factor 
 \be
 \prod_{i=1}^{n/2} \frac{\tau x_i/2}{\sinh( \tau x_i/2)}
 \ee
is anything more than  a mere coincidence. The question of how the $T^2$ contributions to the currents are linked the gravitational anomaly is also raised in \cite{ideal} but was  left unanswered because they work only in flat space. We believe that the illustrative examples in  our section \ref{SEC:Anomaly-derivations} go some of the way to explaining that the similarity  is not a coincidence.
The essential idea is  that when we generate our  temperature from the 1+1 dimensional Schwarzschild sausage we need only to curve  together the  radial  and time  dimensions. As a consequence  only  one   $x_i$ is non-zero,  so only one non-trivial factor  appears  in the A-roof generating function. This also means that when expressed in terms of the Pontryagin classes  only one of the ${\mathfrak p}_n$ can be  non-zero. This will be   ${\mathfrak p}_1(R)=x_1^2 =-T^2$, where  the minus sign accounts for the   difference between $\sinh \tau x/2$ and $\sin \tau T/2$.

\section{Discussion}
\label{SEC:discussion}

In 1967 Sutherland \cite{sutherland} and Veltman   \cite{veltman} argued   that   PCAC and current algebra require  the  decay $\pi_0\to \gamma\gamma$ to be  strongly suppressed ---  a result contrary both to the experimental fact that this is principal decay mode of the neutral pion, and to the fact that the observed decay rate  had been  accurately calculated by Steinberger  in 1949 \cite{steinberger} from a  Pauli-Villars regulated ${\rm AVV}$ triangle diagram. 
Two years later  the  contradiction  was resolved  by Adler \cite{adler} and by Bell and Jackiw \cite{bell-jackiw} who  showed  that the Sutherland-Veltman argument fails    because  it  requires an illegitimate  shift of integration variable in the triangle diagram,  which is only conditionally convergent. As  a consequence  they found that even for massless fermions the  axial current is  not conserved. 

The early understanding  of such  an  anomalous failure   of conservation laws  was mostly  of a formal mathematical  character. The subtle issue of conditionally-convergent Feynman integrals was  followed by   Kiskis, Nielsen and Schroer and others making a   connection with  the mathematically deep Atiyah-Singer index theorem  \cite{kiskis,nielsen-schroer}.    Fujikawa   \cite{fujikawa}  then  showed that  the index-theorem mandated   difference between the number of left-  and right-handed Dirac eigenmodes  led to  the  path-integral measure failing to be  invariant under chiral transformations. 
It was only around  1982 that Peskin \cite{peskin} and others  realized that  in the massless case  the {\it physical\/}  source of the the ${\bf E}\cdot {\bf B}/4\pi^2 $ chiral  anomaly is that  the $|{\bf B}|/2\pi$ density of gapless modes   in the   ${\bf B}$ field  allows a     steady $\dot N= \dot  { k}_\parallel/2\pi = {E}_\parallel/2\pi $ flow of eigenstates  out of the  infinitely deep   Dirac sea, which is   acting  as  a  Hilbert hotel.  At about the same time  Nielsen and Ninomiya \cite{nielsen-ninomiya}   showed that in crystals, where there are necessarily equal numbers of left- and right-handed Weyl nodes, the Hilbert-hotel picture is not needed because the  Dirac seas of left and right-handed fermions pass eigenstates  to one another at their common   seabed. This ambichiral traffic is  the  basis for our present understanding of Weyl semimetals. Later Callan  and Harvey  \cite{callan-harvey}   showed  that, in the case of an uncanceled net  anomaly,   charge is supplied to the bottom of the  Dirac sea {\it via} inflow  from higher dimensions. Their  bulk-edge and bulk-surface connection  is central to the physics of the quantum Hall effect and topological insulators. In the latter  the picture is particularly clear because top and bottom of the branches of gapless boundary-modes  merge with, and emerge from, the  lower  and upper   edges of the higher-dimensional  bulk states'   energy gap.  It is  now also  understood   \cite{son,stephanov} that the spectral flow of eigenstates  can  be be computed by including a Berry-phase induced anomalous velocity  in semiclassical dynamics.
 
Today we have a good {\it mathematical\/}  understanding of gravitational anomalies \cite{alvarez-gaume-witten,alvarez-gaume-ginsparg}, but  a  comparable  {\it physical\/} explanation, analogous to the ${\bf E}\cdot {\bf B}$ spectral flow mechanism, does  not seem to exist. In \cite{stone-dwivedi-zhou1}    an attempt was made to generalize the semiclassical Berry-phase picture to motion in curved space, but the generalization  was frustrated by the unusual Lorentz transformation properties of massless   particles with spin \cite{stephanov-lorentz,stone-dwivedi-zhou2,horvathy1,horvathy2}.  While the main result of the present work is the explicit derivation of the $T^2$ contribution to the CVE from the anomaly,  we hope the simple {\it gedanken\/}  spacetime that we have constructed to do this will be useful  for developing   a physical understanding in the gravitational  case also.

\section{acknowledgements} MS would like to thank Andrzej Woszczyna of Krak{\'o}w Jagiellonian University for his e-mail assistance in  learning how to use  the ccgrg package for Mathematica$\!^{\rm TM}$. We also benefited  from useful comments on the text  by  Karl Landsteiner and Vatsal Dwivedi. This work was not directly supported by any funding agency, but it would not have been possible without  resources provided by the Department of Physics at the University of Illinois at Urbana-Champaign.

\end{document}